\documentclass[amsmath,amssymb, aps, prb, twocolumn, notitlepage]{revtex4-1}

\usepackage{graphicx}
\usepackage{dcolumn}
\usepackage{bm}
\usepackage[usenames]{color}
\usepackage{slashed}
\usepackage[utf8]{inputenc}
\usepackage[caption=false]{subfig}
\begin{document}

\title{Optical Signatures of the Chiral Anomaly in Mirror-Symmetric Weyl Semimetals}

\author{Aaron Hui}
\affiliation{School of Applied \& Engineering Physics, Cornell University, Ithaca, New York 14853, USA}

\author{Yi Zhang}
\email{frankzhangyi@gmail.com}
\affiliation{International Center for Quantum Materials, Peking University, Beijing, 100871, China}
\affiliation{Department of Physics, Cornell University, Ithaca, New York 14853, USA}

\author{Eun-Ah Kim}
\affiliation{Department of Physics, Cornell University, Ithaca, New York 14853, USA}
\date{\today}

\begin{abstract}
    The chiral anomaly is a characteristic phenomenon of Weyl fermions, which has condensed matter realizations in Weyl semimetals. Efforts to observe smoking gun signatures of the chiral anomaly in Weyl semimetals have mostly focused on a negative longitudinal magnetoresistance in electronic transport. Unfortunately, disentangling the chiral anomaly contribution in transport or optical measurements has proven non-trivial. Recent works have proposed an alternative approach of probing pseudoscalar phonon dynamics for signatures of the chiral anomaly in non-mirror-symmetric crystals. Here, we show that such phonon signatures can be extended to scalar phonon modes and mirror-symmetric crystals, broadening the pool of candidate materials. We show that the presence of the background magnetic field can break mirror symmetry strongly enough to yield observable signatures of the chiral anomaly. Specifically for mirror-symmetric Weyl semimetals such as TaAs and NbAs, including the Zeeman interaction at $|\mathbf{B}| \approx 10$T, we predict that an IR reflectivity peak will develop with an $\mathbf{E}_\text{IR}\cdot\mathbf{B}$ dependence.
\end{abstract}

\maketitle

\section{Introduction}

The Weyl semimetal has been generating excitement as a new experimentally realizable class of topological materials in three dimensions.\cite{Yan2017, Armitage2018} 
The materials are so named due to the existence of Weyl points in the momentum space, where two non-degenerate bands intersect and disperse linearly. 
Weyl points are monopoles of Berry curvature and characterized by their chirality, a topological invariant describing the parallel/anti-parallel (right/left-handed) locking between their momentum and spin or pseudo-spin. 
One of the exciting phenomena predicted in the Weyl semimetal is the condensed matter realization of the chiral anomaly: the chiral charge - the population difference between the left and right-handed Weyl fermions - is not conserved after quantization.

The non-conservation of chiral charge means that, under the application of parallel $\mathbf{E}$ and $\mathbf{B}$ fields, particles will be pumped between left-handed and right-handed Weyl points.
Therefore, in the presence of a chiral anomaly, one can think of the $\mathbf{B}$-field as creating a topologically protected channel of charge between left and right-handed Weyl points, whose conductivity and direction are controlled by the magnetic field. 
The presence of this channel leads to the so-called chiral magnetic effect,\cite{Nielsen1983, Kharzeev2008, Fukushima2008, Son2012, Zyuzin2012} where a current will develop along the magnetic field in the presence of a chemical potential difference between Weyl nodes with opposite chirality.
In order to balance the charge transfer, scattering between Weyl nodes is required; this scattering process is rare because the Weyl nodes are generically well-separated, so this conduction channel has high conductivity.
In the limit of large $\mathbf{B}$, intra-node scattering is suppressed within each chiral Landau level, consisting only of a single linear branch. 
The inter-node scattering time, which is longer than the $\mathbf{B}=0$ intra-node scattering time, then controls the conductivity in this limit. 
Therefore, the chiral anomaly leads to a $\mathbf{B}$-field dependent enhancement in the conductivity.\cite{Nielsen1983}
Negative longitudinal magnetoresistance was therefore proposed as a signature of the chiral anomaly in Weyl semimetals.\cite{Son2013, Burkov2014, Burkov2015} 

Indeed, negative magnetoresistance has been observed in a number of Weyl semimetals;\cite{Liang2014, Huang2015, Xiong2015, Zhang2016, LiHui2016, LiQiang2016, Hirschberger2016, Kuroda2017, Niemann2017, Zhang2017, Liang2018} however, negative magnetoresistance was not unique to Weyl semimetals and could potentially be caused by other effects.\cite{Kim2013, He2013, Goswami2015, Wiedmann2016, Dai2017, Andreev2018}
For instance, negative magnetoresistance was also measured in the non-Weyl semimetal materials PdCoO$_2$, PtCoO$_2$, SrRuO$_4$, and Bi$_2$Se$_3$.\cite{Kikugawa2016, Wiedmann2016} 
To complicate matters further, the point contacts used for magnetoresistance measurements were susceptible to current jetting, where the current is focused by a magnetic field, artificially enhancing the measured conductivity and potentially overwhelming the chiral anomaly signature.\cite{Reis2016, Arnold2016}
For these reasons, the chiral anomaly interpretation of electronic transport results has been controversial.

In search of sharper signatures of the chiral anomaly and Weyl semimetals, a number of proposals have been put forth.\cite{Lv2013, Panfilov2014, Zhou2015, Redell2016, Liu2013, Hutasoit2014, Grushin2016, Cortijo2015, Cortijo2016, Pikulin2016, Spivak2016, Jiang2015, Jiang2016} In this paper, we will be particularly interested in phonon-induced optical signatures. 
Through an axial (chirality-dependent) electron-phonon coupling, a phonon can induce a dynamical chemical potential difference between Weyl points with opposite chirality, which in turn gives rise to a dynamical realization of the chiral anomaly. 
Recent works have found that this can result in anomalous optical features in IR and Raman spectroscopy.\cite{Ashby2014, Song2016, Rinkel2017, Rinkel2019}
However, based on symmetry considerations, it was argued that a phonon mode in a 1D representation can only have an axial coupling if it is pseudoscalar (changes sign under improper rotations).\cite{Song2016} 
As the allowed phonon modes are constrained by the crystal symmetry, pseudoscalar phonons only exist in crystals where the mirror symmetries are sufficiently broken.\cite{Song2016} 
Therefore, previous works ruled out such chiral-anomaly induced optical phenomena in Weyl semimetal candidates with many mirror planes, such as TaAs and NbAs.\cite{Song2016, Rinkel2017}

We claim, by contrast, that such optical signatures of the chiral anomaly can occur in all mirror-symmetric crystals for both scalar and pseudoscalar phonons, due to the role of a necessary magnetic field. 
Previous analyses\cite{Song2016, Rinkel2017} assumed the Weyl points to be locally identical (up to chirality) and the linear dispersion to be isotropic. 
If one breaks these assumptions and allows the Fermi velocities to differ, a scalar phonon can also develop an effective, non-vanishing axial coupling.
Such a difference in Fermi velocities can be induced by the magnetic field necessarily present in the experiments.
Because of this, it is important to consider the effect of magnetic field on symmetries neglected in previous analyses.

The magnetic field, a pseudovector, changes sign under improper rotation; under the reflection $x\rightarrow -x$, the magnetic field transforms as $\left(B_x,B_y,B_z\right) \rightarrow \left(B_x,-B_y,-B_z\right)$. 
Therefore, it breaks all mirror symmetries except for the mirror plane normal to it, if such a mirror plane exists. 
The Zeeman effect and the Landau level quantization are examples of such mirror-symmetry-breaking effects. 
In the presence of at most one mirror plane, an effective pseudoscalar phonon is allowed to exist, so the axial component of the phonon coupling for this mode is generically non-zero. 
Since optical signatures of the chiral anomaly require the presence of a static magnetic field, no symmetry restrictions on Weyl semimetals are required to see this signature. In this paper, by considering a suitable microscopic model, we show that the Zeeman effect and the Landau level quantization can result in substantial Fermi velocity asymmetry that can drive detectable optical signatures of chiral anomaly.

The outline of the paper is as follows: In Section II, we introduce a tight-binding model Hamiltonian in the same symmetry class as TaAs and NbAs and analyze the effect of mirror-symmetry-breaking Zeeman effect and Landau level quantization on the fermion dynamics. 
In Section III, we discuss the electron-phonon coupling and its symmetry constraints for optical signatures. 
In Section IV,  given the magnetic field's mirror-asymmetric effect on the Fermi velocities, we estimate the strength and visibility of the IR reflectivity signal corresponding to the dynamically-driven chiral anomaly. 
Finally, we conclude our results and discuss their distinction from multiferroic materials in section V.

\section{Tight-binding model of 3D Weyl fermions with magnetic field}

To quantitatively analyze the symmetry-breaking effect of the magnetic field, we consider the following 3D electronic tight-binding model with crystal symmetries identical to the Weyl semimetals TaAs and NbAs:\cite{Ramshaw2018}
\begin{align}
    H_{0} = &t\sum_{\langle ij\rangle,s} c^\dagger_{is}c_{js} + \sum_{i, s}\Delta_i c^\dagger_{is} c_{is} \nonumber\\ 
    &+ i\lambda\sum_{\langle\langle ik \rangle\rangle,ss'} c^\dagger_{is}c_{ks'} \sum_{j}\mathbf{d}_{ijk}\cdot\bm{\sigma}_{ss'} \label{eq:zero-field-model}
\end{align}
where $t$ is the nearest neighbor hopping, $\Delta_i = \pm \Delta$ is a staggered potential whose sign depends on the sublattice being a Ta(Nb) or As site, and $\lambda$ is the amplitude of the spin-orbit interaction between next-nearest neighbors. $s=\uparrow,\downarrow$ denotes spin, and $\bm{\sigma}$ are the Pauli matrices. The vector $\mathbf{d}_{ijk} = \mathbf{d}_{ij}\times \mathbf{d}_{jk}$, where $j$ is an intermediate site between $i$ and $k$, and $\mathbf{d}_{ij}$ is the displacement vector from $i$ to $j$. 

In the absence of the magnetic field, the model is time-reversal invariant and breaks inversion symmetry. 
Two mirror planes exist in the $xz$ and $yz$ directions. 
For large values of $\lambda$, the model is a 3D topological insulator; for large values of $\Delta$, on the other hand, the model is a normal insulator. 
In between, a time-reversal-invariant Weyl semimetal exists in a finite phase space, for instance, at $t=500$meV, $\Delta=350$meV, $\lambda=100$meV; we will use these parameters throughout this paper.
Comparing this model at $B=0$T to DFT calculations of the TaAs band structure\cite{Huang2015, Yan2017, Ma2017} and the measured Fermi velocities around the Weyl points,\cite{Ramshaw2018} we find qualitative agreement. More details on the low-energy electronic properties of the model can be found in the Appendix. 

In the presence of a magnetic field, we generally expect the Hamiltonian to change in two ways.
One modification is the Zeeman effect, describing the coupling of the electron spin to the magnetic field given by
\begin{align}
    H_{z} = g\mu_B \sum_{iss'} c^\dagger_{is}c_{is'} \mathbf{B}\cdot\bm{\sigma}_{ss'} \label{eq:zeeman}
\end{align}
with $g$ the $g$-factor, $\mu_B$ the Bohr magneton, and $\mathbf{B}$ the magnetic field.
We estimate a large $g$-factor $g\approx 50$ for typical topological Weyl semimetal materials with strong spin-orbit coupling, such as TaAs and NbAs, by analogy to measurements in related materials.\cite{Singh1982, Hu2016}
The inclusion of the Zeeman effect at finite $\mathbf{B}$ breaks the time-reversal symmetry and all mirror plane symmetries except the mirror plane normal to the magnetic field, if it exists. 

The other modification, which we refer to as the Landau level quantization, comes from the minimal coupling of the electromagnetic vector potential to the electron current. To incorporate this effect, we perform the Peierls substitution on the kinetic term and the spin-orbit interaction:
\begin{align}
    &c^\dagger_{is}c_{js} \rightarrow e^{i A_{ij}} c^\dagger_{is}c_{js} \nonumber\\ 
    &c^\dagger_{is}c_{ks'} \rightarrow e^{i A_{ik}} c^\dagger_{is}c_{ks'}
    \label{eq:LLquantization}
\end{align}
where $A_{ij}$ and $A_{ik}$ are the electromagnetic vector potentials (integrated) from $i$ to $j$ and from $i$ to $k$, respectively. We've chosen to set the electron charge $e = 1$ (and $\hbar = 1$, as usual). 
We also use the lattice constants of TaAs to convert the magnetic flux into the magnetic field in unit of Tesla. 
As is well known, minimal coupling to a magnetic field leads to a quantization of the electronic dispersion into separate Landau bands.
In particular, the dispersion normal to the magnetic field becomes quantized, so the dispersion becomes one-dimensional with bandgap controlled by the magnetic field.
Similar to the Zeeman effect, Landau level quantization also breaks time-reversal symmetry and all mirror plane symmetries except the (possibly existent) mirror plane normal to the magnetic field.

\begin{figure}
    \centering
    \includegraphics[width=.98\linewidth]{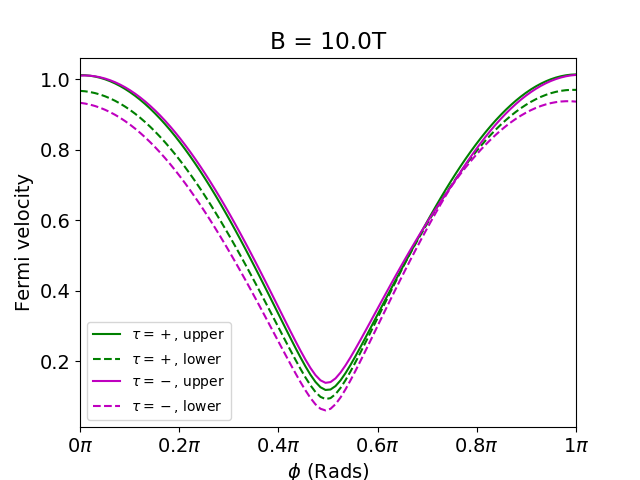}
    \caption{The magnitude of the Fermi velocity as a function of the azimuthal angle $\phi$ in the $k_x-k_y$ plane for a pair of Weyl points, denoted in green and magenta, originally related by the mirror symmetry at $|\mathbf{B}|=0T$. The solid and dashed lines denote the upper and lower branches of the Weyl dispersion, respectively. For a magnetic field $|\mathbf{B}|\sim 10$T in the $\hat x/2 + \sqrt{3}\hat y /2$ direction, the differences developed between these curves demonstrate the mirror-symmetry breaking of the Zeeman effect. }
    \label{fig:fermivelocity}
\end{figure}

Let us focus on the impact of magnetic-field-induced mirror symmetry breaking on the low-energy dispersion of the Weyl nodes near the $k_z \approx 0$ plane in the Brillouin zone. For clarity, we will consider the effects of the Zeeman effect (Eq. \ref{eq:zero-field-model} and \ref{eq:zeeman}) and the Landau level quantization (Eq. \ref{eq:zero-field-model} and \ref{eq:LLquantization}) separately.

For the Zeeman interaction, we diagonalize the Hamiltonian $H_0+H_z$ in $\vec k$ space as Eq. \ref{eq:zeeman} preserves lattice translation symmetries. We find that even with a magnetic field as large as $|\mathbf{B}|=10T$, the Weyl nodes only displace a scale $\sim 0.1\%$ of the Brillouin zone (see the Appendix). 
Therefore, the impact of the Zeeman effect due to the $k$-dependence of the electron-phonon coupling is likely small, and we neglect this contribution.  
On the other hand, the symmetry breaking from the magnetic field has a more prominent effect on the Fermi velocities, especially in topological semimetal models and materials with strong spin-orbit interactions, so that the Zeeman spin-splitting effect strongly impacts electron velocity.
In Fig.~\ref{fig:fermivelocity}, we see that the Fermi velocities of the Weyl points connected via mirror symmetries initially identical at zero field clearly become different when a magnetic field is turned on. 

\begin{figure}
    \centering
    \includegraphics[width=1.05\linewidth]{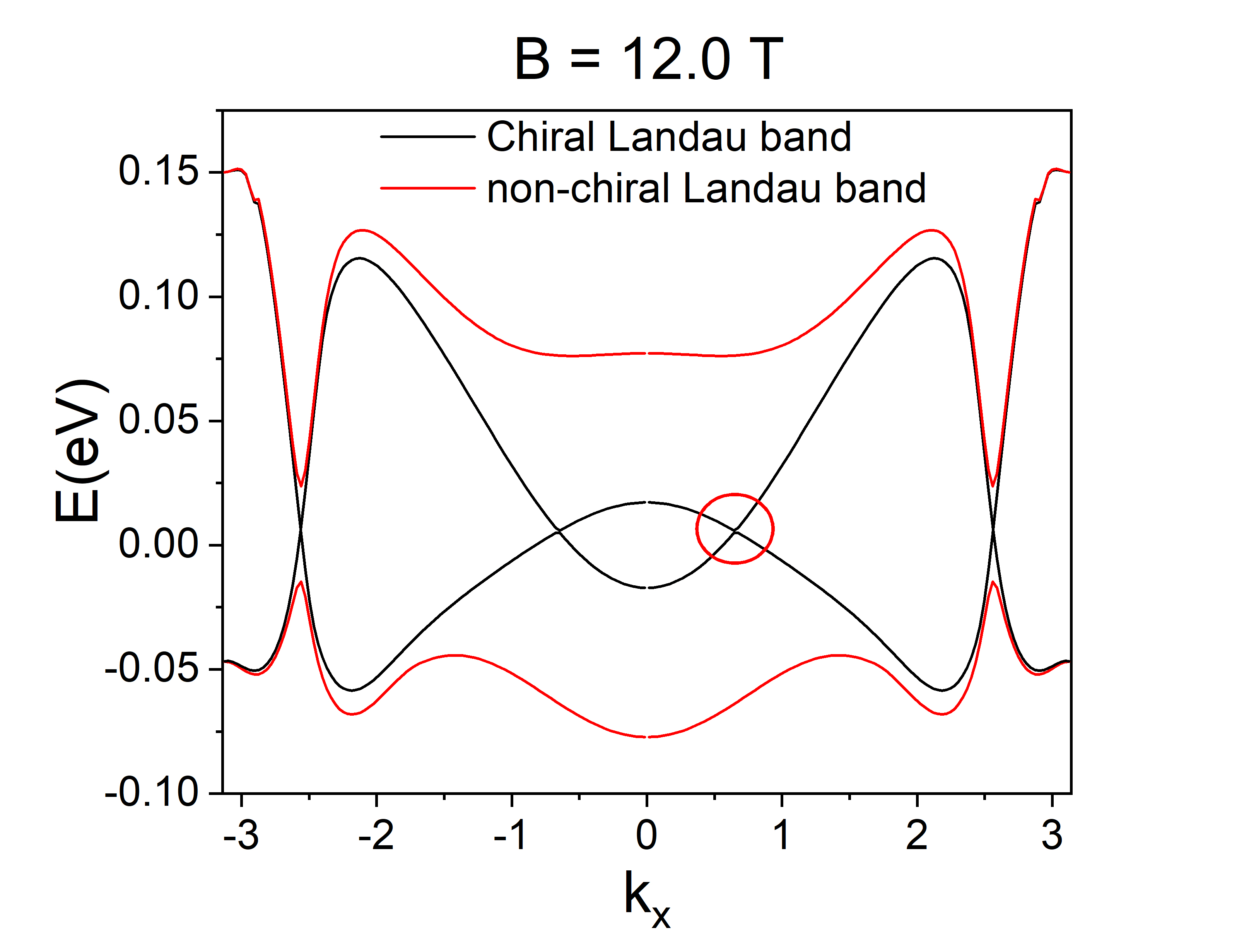}
    \caption{The $k_x$ dispersion of the four Landau bands closest to the Weyl node energy in the presence of the Landau level quantization of a magnetic field $|\mathbf{B}|\sim 12$T in the $\hat x$ direction. The eight gapless linear branches are the chiral Landau bands descending from the eight Weyl nodes, respectively, and responsible for the electronic properties at low energy. A finite (indirect) gap separates the other Landau bands. As an example, the chiral Landau bands in the red circle as the descendants of a pair of Weyl nodes are illustrated in Fig. \ref{fig:LLBreaking}. }
 \label{fig:1dkxdisp}
\end{figure}

\begin{figure}
    \centering
    \includegraphics[width=.98\linewidth]{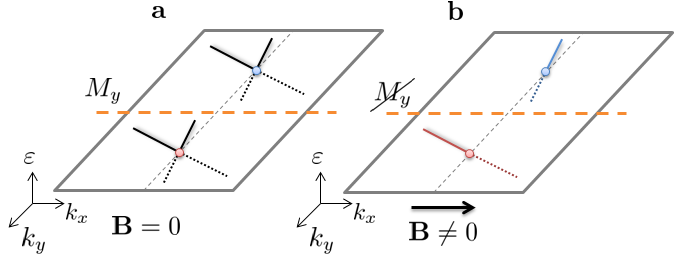}
    \caption{(a) A schematic plot of the $k_x$ dispersion of a pair of Weyl nodes of opposite chirality (labeled with blue and red) related by a $M_y$ mirror symmetry at zero field, and (b) the chiral and anti-chiral Landau bands selected out in the presence of a magnetic field along the $x$-direction. Since the chiral and anti-chiral Landau levels can generically have distinct Fermi velocities, they explicitly break the $M_y$ symmetries and contribute to an effective axial electron-phonon coupling.}
 \label{fig:LLBreaking}
\end{figure}

For Landau level quantization, we focus our attention on the linear, chiral Landau bands. 
We specialize to $\mathbf{B} = B \hat{x}$ for simplicity and introduce the electromagnetic vector potential via Eq.~\eqref{eq:LLquantization}. 
Consequently, the dispersion along $k_y$ and $k_z$ becomes quenched, and the discrete Landau bands disperse only along $k_x$, which remains as a good quantum number. 
Using exact diagonalization for the Hamiltonian within a magnetic unit cell, we obtain the one-dimensional $k_x$ dispersion - see Fig. \ref{fig:1dkxdisp} for an example at $|\mathbf{B}|\sim 12$T. 
It is important to note that the branch of the dispersion that evolves into the chiral Landau band depends on the chirality of each Weyl node,\cite{Nielsen1983} schematically shown in Fig.~\ref{fig:LLBreaking}. 
Therefore, despite the identical zero-field dispersion of a pair of mirror-symmetric Weyl nodes, the differing chiralities ensure that the chiral and anti-chiral Landau bands selected out by the magnetic field generally have distinct Fermi velocities. 
Interestingly, such a difference between the Fermi velocity of the chiral and anti-chiral Landau bands is a form of mirror-symmetry breaking, depending on the anisotropy of the original Weyl fermions instead of the strength of the magnetic field. 
The remainder of the Landau bands will be gapped by the magnetic field, so that the chiral branches dominate near the Fermi energy, see Fig. \ref{fig:1dkxdisp}.
As a result, chiral anomaly effects may become visible if the cyclotron energy of $\mathbf{B}$ is sufficiently large and the Fermi energy sufficiently close to the Weyl node.

In summary, the mirror symmetry connecting a pair of Weyl nodes is explicitly broken by a magnetic field. 
The magnetic-field-induced difference between the Weyl nodes' Fermi velocities, induced by the Zeeman effect and the chiral selectivity of Landau level quantization, are physical manifestations of the broken mirror symmetry.
We will discuss its phenomenological consequences for the dynamical chiral anomaly in Sec. IV. 

\section{Electron-phonon Coupling and Symmetry Constraints}

To understand the impact of the Weyl fermion dynamics and its symmetry constraints on the electron-phonon coupling, we consider the interaction between phonons and a pair of Weyl nodes with opposite chirality $\tau =\pm 1$:
\begin{align}
    \mathcal{H}_\text{ep} = \sum_{\mathbf{kq}}\sum_{\sigma\sigma'\tau}\left(\sum_\lambda u_{\sigma\sigma',\tau}^\lambda(\mathbf{q}) v_{\mathbf{q}\lambda}\right) c^\dagger_{\mathbf{k}\sigma\tau} c_{\mathbf{k-q}\sigma'\tau} \label{eq:ep-coupling}
\end{align}
where $v_{\mathbf{q}\lambda}$ is the phonon displacement operator in mode $\lambda$ at momentum $\mathbf{q}$ and $\sigma,\sigma'$ describe the pseudospin of the electrons. We have neglected inter-node electron scattering, since it requires a large momentum transfer $q$ to connect the well-separated Weyl nodes in the momentum space. 
Decomposing the electron-phonon coupling into its irreducible representations,
\begin{align}
    u^\lambda_{\sigma\sigma',\tau} = u^\lambda_{00}\delta_{\sigma\sigma'} + \mathbf{u}_0^\lambda\cdot\bm{\sigma}_{\sigma\sigma'} + \tau(u_{0z}^\lambda\delta_{\sigma\sigma'} + \mathbf{u}_z^\lambda\cdot\bm{\sigma}_{\sigma\sigma'})
\end{align}
The two latter terms correspond to the (chirality-dependent) axial coupling responsible for the chiral anomaly. We focus on the axial coupling constant $u_{0z}^\lambda$ since the contribution from $\mathbf{u}^\lambda_z$ is suppressed by a factor of $v_{\tau}/c$, as we will see later.

The symmetries of the system impose constraints on the electron-phonon coupling. 
In particular, $\mathbf{u}_z^\lambda$ vanishes in the presence of time-reversal symmetry, while $u_{0z}^\lambda$ vanishes in the presence of two non-coplanar mirror-symmetry planes.\cite{Song2016, Rinkel2017, Rinkel2019} Therefore, it seems that the mirror symmetry in the crystal should be sufficiently broken to host a nontrivial phonon signature as a result of the chiral anomaly. We find, on the other hand, that the imposed magnetic field can break the mirror symmetries sufficiently for the signatures to appear in a much broader pool of Weyl semimetal candidates.

For our tight-binding model in Eq. \eqref{eq:zero-field-model}, we expect the magnetic-field-induced changes to $u^\lambda_{\sigma\sigma',\tau}$ due to the small displacements of the Weyl point locations to be sub-dominant; instead, the key ingredient that leads to interesting phonon behavior is the induced change in Fermi velocity, which we discuss next. 

\section{Estimating the Effect of Magnetic field on the Fermi Velocity}
\label{Estimation}

In this section, we study the chiral anomaly contribution to the phonon dynamics by integrating out the electronic degrees of freedom.
The low-energy effective theory of our tight-binding model, described by Eqs.~(\ref{eq:zero-field-model}-\ref{eq:LLquantization}), can be captured by the following single-particle Hamiltonian  
\begin{align}
    \mathcal{H}_{\tau} = v_{\tau}(\hat{k})\tau\bm{\sigma}\cdot(-i\bm{\nabla} + e\mathbf{A}) - eA_0
    \label{eq:linearWeyl}
\end{align}
which describes a Weyl point with chirality $\tau =\pm 1$ and anisotropic Fermi velocity $v_{\tau}(\hat{k})$. The terms $A_0,\mathbf{A}$ are the electromagnetic vector potential.\footnote{We assume that the energy separation of the Weyl nodes is zero; the momentum separation is large and presumably irrelevant in the low-energy theory, so it has been dropped.}
Because phonons do not couple electrons between Weyl nodes, the integration over electronic degrees of freedom factorizes between Weyl points (at the leading order); we can restrict our attention to a single pair.

For a pair of Weyl nodes with isotropic and identical Fermi velocity $v_{\tau}(\hat{k}) = v_F$, on integrating out the fermions one finds that the chiral anomaly contributes to a mode-effective phonon charge $\delta\mathbf{Q}$, and hence to a dielectric susceptibility $\chi$:\cite{Rinkel2017}
\begin{align}
    \delta\mathbf{Q}_{-\mathbf{q}\lambda}(-q_0) &= i \frac{e^2\mathcal{V}_c \sqrt{N}}{\pi^2\hbar^2}\frac{\mathbf{B}}{q^2}(q_0 u_{0z}^\lambda - v_F\mathbf{q}\cdot\mathbf{u}_0^\lambda) \label{eq:phononcharge}\\
    \chi^\lambda_{jj'}(q_0, \mathbf{q}) &=  \frac{1}{M\mathcal{V}_c}\frac{\delta Q_{\mathbf{q}\lambda j} \delta Q_{\mathbf{q}\lambda j'}}{\omega^2_{\mathbf{q}\lambda}+i\kappa u_{00}^\lambda \mathbf{q}\cdot\delta\mathbf{Q}_{\mathbf{q}\lambda}-q_0^2} \label{eq:dielectric}
\end{align}
where $(q_0,\mathbf{q})$ is the frequency-momentum vector of the phonon, $\mathcal{V}_c$ is the unit cell volume, $M$ is the total mass of ions in the unit cell, $N$ is the number of unit cells, and $\mathbf{B}$ is the static background magnetic field. $\kappa = \sqrt{N}/(Me)$, $q^2 = q_0^2 - v_F^2 \mathbf{q}^2$, and $\omega_{\mathbf{q}\lambda}$ is the bare phonon dispersion of mode $\lambda$. 
Since $q_0 = c\mathbf{q}$ for light, the $\mathbf{u}_0^\lambda$ term is suppressed by $v_F/c$. 
When the IR light is on resonance with the phonon driving the chiral anomaly, the dielectric constant diverges and the reflectivity develops a peak with a form factor $\mathbf{E}_\text{IR}\cdot\mathbf{B}$. Also, such chiral anomaly contribution to $\chi_{jj'}^\lambda$ clearly depends on a non-zero axial coupling constant $u_{0z}^\lambda$.

\begin{figure}
    \centering
    \includegraphics[width=\linewidth]{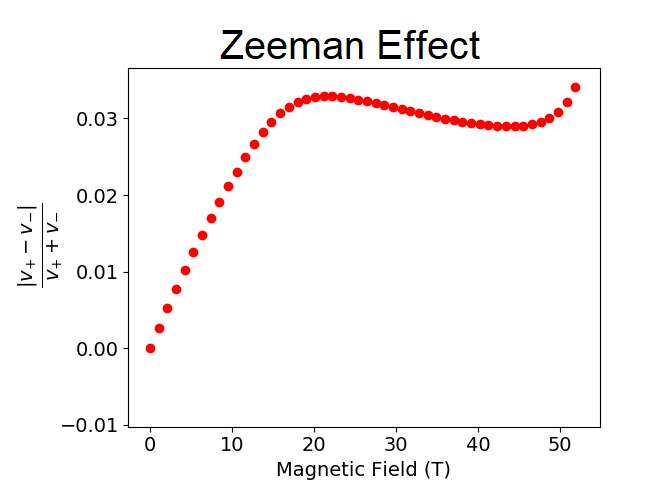}
    \caption{The relative difference (induced by the Zeeman effect) between the Fermi velocities of a pair of Weyl nodes as a function of the strength of the magnetic field $\mathbf{B}$ in the $\hat x/2 +\sqrt{3}\hat y/2$ direction. The ratio is averaged over all directions. Assuming that $u_{00}^\lambda$ is the only non-zero electron-phonon coupling component at $\mathbf{B}=0$, this quantity measures the ratio $u_{0z}^\lambda/u_{00}^\lambda$ generated by the inclusion of the magnetic field and the broken symmetry between $v_+$ and $v_-$ (see Eqs.~\eqref{eq:u00-renorm} and \eqref{eq:u0z-renorm}).}
    \label{fig:velocitypercent}
\end{figure}

\begin{figure}
    \centering
    \includegraphics[width=\linewidth]{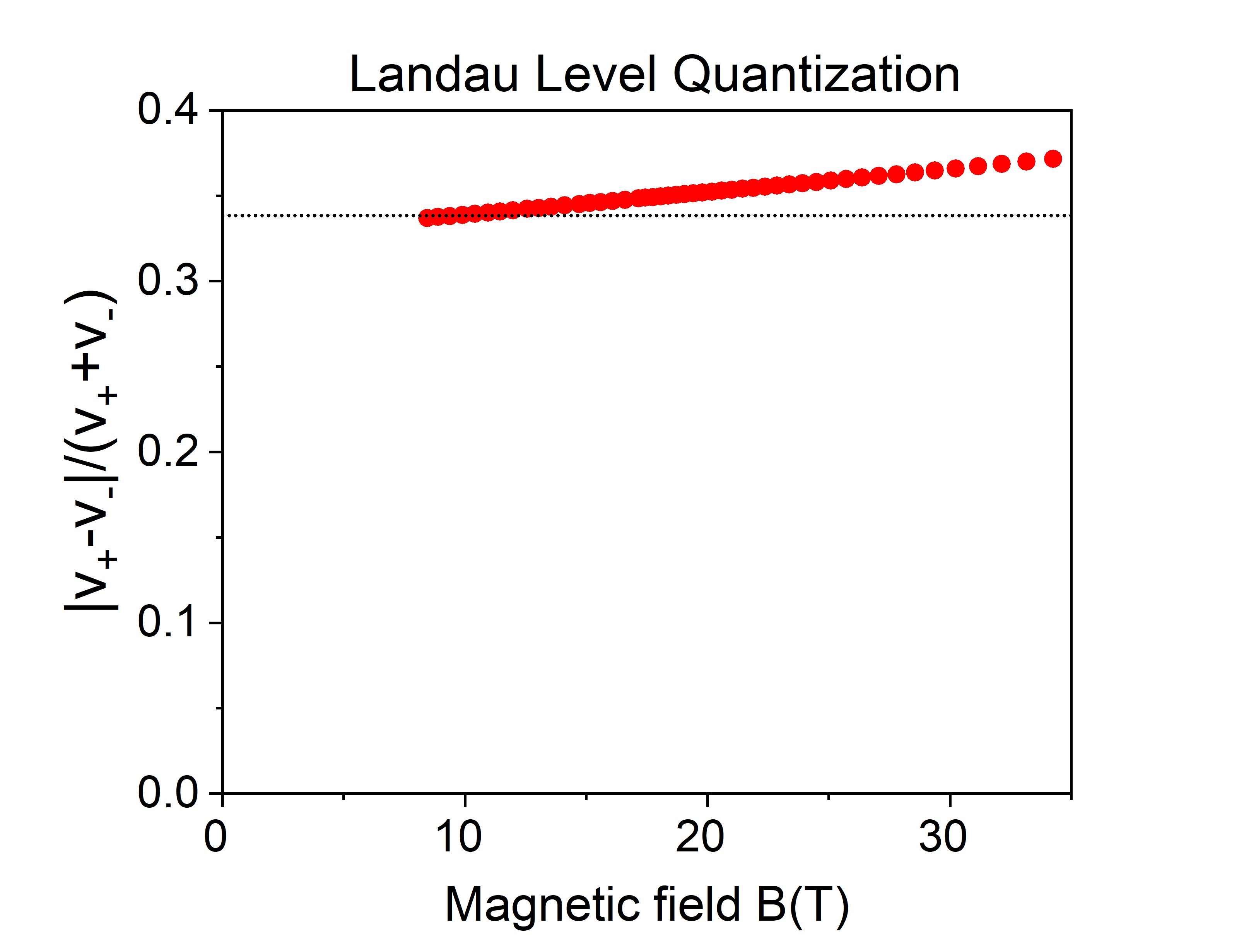}
    \caption{The relative difference (induced by the Landau level quantization) between the chiral Landau level Fermi velocities descending from a pair of Weyl nodes as a function of the magnetic field $\mathbf{B}$ along the $\hat x$ direction. Similar to Fig. \ref{fig:velocitypercent}, the value of $\left|v_+-v_-\right|/\left(v_+-v_-\right)$ measures the ratio $u_{0z}^\lambda/u_{00}^\lambda$ generated by the magnetic field. The black dotted line is the value evaluated with the zero-field dispersion.}
    \label{fig:velocityvsB}
\end{figure}

In comparison, our generalized model in Eq.~\eqref{eq:linearWeyl} takes into account the anisotropic Fermi velocity around a Weyl node as well as the different Fermi velocities between the Weyl nodes. We consider a totally-symmetric scalar phonon mode at zero field, where all components of the electron-phonon coupling are $0$ except $u_{00}^\lambda$.
For simplicity, this system can be mapped back to the isotropic case by rescaling the fermions by $v_{\tau} c^\dagger_{\tau} c_{\tau} \rightarrow v_F c^\dagger_{\tau} c_{\tau}$, which changes the electron-phonon coupling and induces components in the non-identity piece:
\begin{align}
    u_{00}^\lambda &\rightarrow \frac{v_F}{2}\left(\frac{1}{v_+} + \frac{1}{v_{-}}\right)u_{00}^\lambda \label{eq:u00-renorm}\\
    u_{0z}^\lambda &\rightarrow \frac{v_F}{2}\left(\frac{1}{v_+} - \frac{1}{v_{-}}\right)u_{00}^\lambda \label{eq:u0z-renorm}
\end{align}
The rescaling of the fermions also changes $A_0$, but it does not affect the phonon charge and dielectric susceptibility in Eqs.~\eqref{eq:phononcharge} and \eqref{eq:dielectric} so we neglect the change.
As is manifest after rescaling, the difference of the Fermi velocity is equivalent to an axial component $u_{0z}^\lambda$ in the isotropic setting since $u_{0z}^\lambda/u_{00}^\lambda = |v_{+}-v_{-}|/(v_{+}+v_{-})$.
For the Zeeman effect, a non-zero difference develops between the Fermi velocities of the pair of Weyl nodes related by the original mirror symmetry. 
The difference is generally greater at larger magnetic field, see Fig.~\ref{fig:velocitypercent}, and $u_{0z}^\lambda \sim 0.02u_{00}^\lambda$ at $10T$ within our model. 
For Landau level quantization, on the other hand, the non-zero difference between $v_{+}$ and $v_{-}$ originates from the anisotropy of the dispersion around each Weyl point. 
Also, the difference is less dependent on $\mathbf{B}$, see Fig. \ref{fig:velocityvsB}, as long as $\mathbf{B}$ is large enough to separate the non-chiral Landau bands and suppress their contribution. 
Landau level quantization gives $u_{0z}^\lambda \sim 0.3u_{00}^\lambda$ within our highly anisotropic model, yet it is also possible that $u_{0z}^\lambda \rightarrow 0$ irrespective of $\mathbf{B}$ when the anisotropy vanishes, e.g. for two isotropic Weyl points.

\begin{figure}
    \centering
    \includegraphics[width=.98\linewidth]{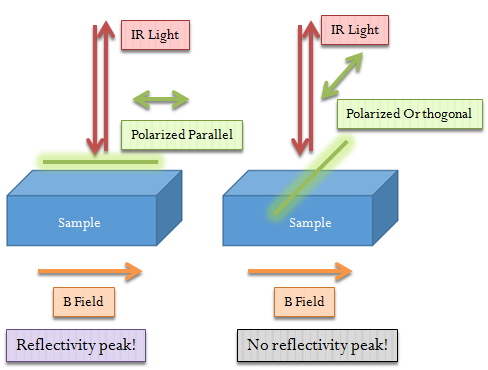}
    \caption{Proposed experimental setup to measure the IR signature of the chiral anomaly. In the presence of collinear $\mathbf{E}_\text{IR}$ and $\mathbf{B}$ fields, a peak in optical reflectivity is expected for inducing pseudoscalar phonon modes that couple strongly to the Weyl fermion electrons. Such effect also displays a $\mathbf{E}_\text{IR}\cdot\mathbf{B}$ dependence as one rotates $\mathbf{E}_\text{IR}$ relative to $\mathbf{B}$ in experiments.}
    \label{fig:experiment}
\end{figure}

Now that we have obtained an estimate for the effective $u_{0z}^\lambda$, let's estimate the strength of the corresponding IR signature. 
For example, we focus on the $A_1$ phonon mode in TaAs.
We take $\omega = 8$ THz to match the experimental observation of an $A_1$ phonon mode in TaAs\cite{Xu2017}, $\mathcal{V}_c = 125\text{\AA}$, and $M = 10^{-25}$kg. 
We also take $u_{0z}^{A_1} \sim 0.02u_{00}^{A_1}$, which is reasonably obtainable given either the Zeeman effect with $g=50$ at $|\mathbf{B}| = 10$T or Landau level quantization with the anisotropy in the NbAs and TaAs Weyl dispersion, as previously demonstrated. 
We also estimate $\sqrt{N}u_{00}^\text{A$_1$} \sim 1\text{Ry}/a_B$ on dimensional grounds\cite{Rinkel2017}, and neglect the $\mathbf{u}_z$ contribution given $v_F \ll c$. 
As a result, we obtain $|\delta \mathbf{Q}| \approx .8e$. Next, we calculate the impact of the chiral anomaly on the susceptibility. 
If we drive the IR frequency at $q_0 = 7.9$ THz, corresponding to a resonance width of 6.7 cm$^{-1}$, we find that $\chi^\text{A$_1$}_{zz} = 60 \epsilon_0$. 
Comparing to the experimentally measured zero-field reflectivity $R = \frac{|1-\sqrt{\epsilon_r}|^2}{|1+\sqrt{\epsilon_r}|^2}$ on TaAs crystals\cite{Xu2017}, the chiral anomaly contribution to the reflectivity should be of sufficient weight to be observable over the background of $\chi \approx 400 \epsilon_0$. 
Therefore, we propose an $\mathbf{E}_\text{IR}\cdot\mathbf{B}$ dependent peak in the IR reflectivity as a signature of the chiral anomaly following the experimental setup in Fig.~\ref{fig:experiment}, even for scalar phonon modes and mirror-symmetric Weyl semimetals.\footnote{Note that the proposed signature is a characteristic of the bulk, hence the incident light must penetrate into the bulk for this effect to manifest.}

\section{Discussions and Conclusions}

In this paper, we have focused on utilizing the mirror-symmetry breaking of the magnetic field to realize dynamical chiral anomaly in mirror-symmetric crystals and exhibit optical signatures for scalar phonons in IR spectroscopy.
We would like to emphasize that so long as a magnetic field is present, at most one mirror symmetry remains, so that the axial phonon coupling $u_{0z}^\lambda$ is generically \emph{allowed} from symmetry considerations and a chiral-anomaly induced IR response should be present.
For the specific case where a single mirror plane remains, a pseudoscalar phonon mode normal to the mirror plane is still allowed\cite{Song2016, Rinkel2019}. 
Since both the effective pseudoscalar phonon and the Weyl fermion chirality change sign under mirror symmetry, the axial component of electron-phonon coupling is not restricted to zero, and the corresponding IR signature of the dynamical chiral anomaly survives.\cite{Song2016, Rinkel2019}

Inducing changes in dielectric susceptibility via a magnetic field is a magnetoelectric effect and not completely new.\cite{Rinkel2017}
However, magnetoelectric effects are typically associated with multiferroic materials (e.g., Cr$_2$O$_3$) and previous studies have focused on linear magnetoelectric effects (e.g., $\mathbf{P} \propto \mathbf{B}$).
For the chiral anomaly, the effect is cubic with a characteristic $\mathbf{E}\cdot\mathbf{B}$ signature (i.e. $\mathbf{P} \propto \mathbf{\left(E\cdot B\right) B}$), and known Weyl semimetals are not multiferroic. 
Therefore, we believe that the chiral-anomaly-activated phonon dynamics and IR signatures should be visible in generic Weyl semimetals. 

{\bf Acknowledgements}
We would like to thank C. Fennie, I. Garate, D. Jena, and B. Ramshaw for helpful discussions and comments.
A.H. was supported by the National Science Foundation Graduate Research Fellowship under Grant No. DGE-1650441. 
Y.Z. was supported by the Bethe fellowship at Cornell University and the startup grant at Peking University. 
E.-A. K. was supported by the National Science Foundation through the Platform for the Accelerated Realization, Analysis, and Discovery of Interface Materials (PARADIM) under Cooperative Agreement No. DMR-1539918.

\bibliography{biblio}

\appendix*
\onecolumngrid
\section{Low-energy Weyl dispersion and Weyl nodes of the tight binding model}

\begin{figure}[h!]
    \centering
    \includegraphics[width=.55\linewidth]{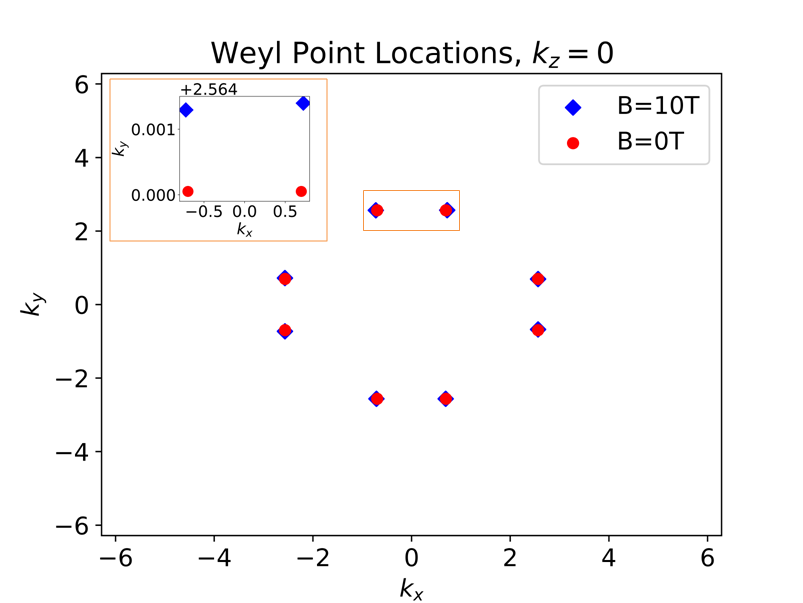}\includegraphics[width=.5\linewidth]{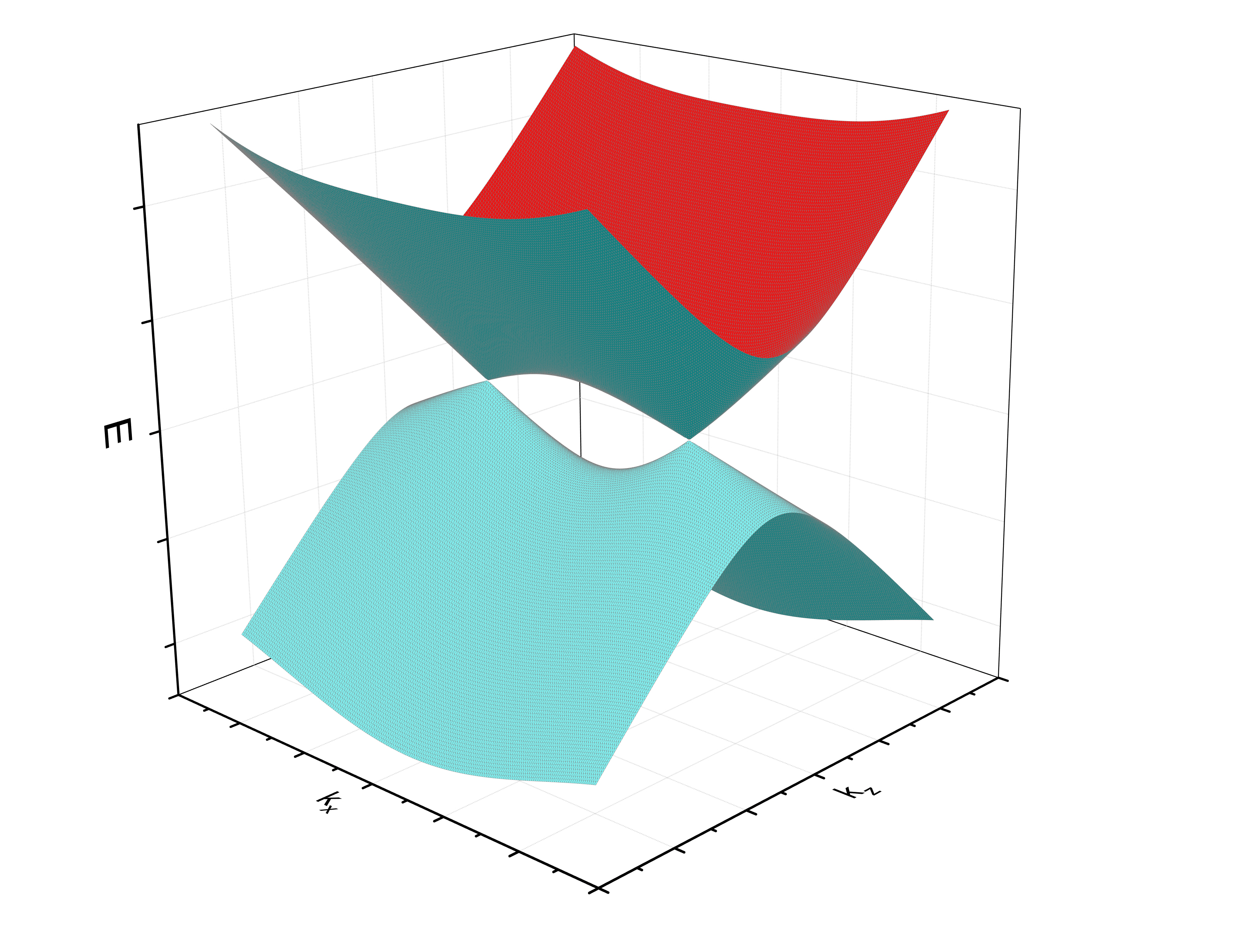}
    \caption{Left: the momentum-space locations of the Weyl nodes on the $k_z=0$ plane show the mirror symmetry is broken in the presence of a magnetic field $|\mathbf{B}|=10$T along the $\hat x/2 + \sqrt{3} \hat y/2$ direction. Note that even for a large $g$-factor $g=50$ and a large magnetic field of $10$T, the Weyl nodes only displaces by a scale $\sim 0.1\%$ of the Brillouin zone. The inset shows a magnified view of the pair of Weyl points in the orange box. Right: The zero-field dispersion in the $k_y-k_z$ plane is approximately linear near the Weyl nodes.}\label{fig:weylptlocs}
\end{figure}

The tight-binding model of Eq. \eqref{eq:zero-field-model} in the main text has four pairs of Weyl nodes on the $k_z=0$ plane at $|\mathbf{B}|=0$T, shown as the red dots in Fig. \ref{fig:weylptlocs} left panel. These Weyl nodes are related to each other by the reflection planes in the $xz$ and $yz$ directions. The low-energy electronic dispersion is approximately linear near each of the Weyl nodes, see Fig. \ref{fig:weylptlocs} right panel.

In the presence of a magnetic field $\mathbf{B}$, these reflection symmetries are generally broken. As a result, the locations of the Weyl nodes are no longer mirror symmetric. However, with the inclusion of the Zeeman effect (Eq.~\eqref{eq:zeeman}), the displacements of the Weyl node locations are relatively small at experimentally relevant parameters, and unlikely to impact the electron-phonon coupling through its $k$-dependence in a meaningful way.

\end{document}